\documentclass[conference]{IEEEtran}

\usepackage{amsmath}
\usepackage[T1]{fontenc}% optional T1 font encoding
\usepackage[utf8]{inputenc}
\usepackage{amssymb,textcomp}
\usepackage{macros}
\usepackage{tikz}
\usepackage{array}
\usepackage{colortbl}
\usepackage{multirow}
\usepackage{textcomp,booktabs,slashbox}

\usetikzlibrary{chains}
\renewcommand{\baselinestretch}{1}
\newcolumntype{P}[1]{>{\centering\arraybackslash}p{#1}}

\begin{document}
\title{R2-D2: Colo\textsl{R}-inspired Convolutional Neu\textsl{R}al Network (CNN)-based Androi\textsl{D} Malware \textsl{D}etections}
\author{TonTon Hsien-De~Huang$^\dagger$,
		Hung-Yu Kao$^*$\\
Leopard Mobile Inc. (Cheetah Mobile Taiwan Agency), Taiwan$^\dagger$\\
Department of Computer Science and Information Engineering, National Cheng Kung University, Taiwan$^\dagger$ $^*$\\
TonTon@TWMAN.ORG$^\dagger$
}

\markboth{Journal of \LaTeX\ Class Files,~Vol.~14, No.~8, August~2015}%
{Shell \MakeLowercase{\textit{et al.}}: Bare Demo of IEEEtran.cls for IEEE Communications Society Journals}

\maketitle

\begin{abstract}
The influence of Deep Learning on image identification and natural language processing has attracted enormous attention globally. The convolution neural network that can learn without prior extraction of features fits well in response to the rapid iteration of Android malware. The traditional solution for detecting Android malware requires continuous learning through pre-extracted features to maintain high performance of identifying the malware. In order to reduce the manpower of feature engineering prior to the condition of not to extract pre-selected features, we have developed a colo\textsl{R}-inspired convolutional neu\textsl{R}al networks (CNN)-based Androi\textsl{D} malware \textsl{D}etection (\textsl{R2-D2}) system. The system can convert the bytecode of classes.dex from Android archive file to rgb color code and store it as a color image with fixed size. The color image is input to the convolutional neural network for automatic feature extraction and training. The data was collected from Jan. 2017 to Aug 2017. During the period of time, we have collected approximately 2 million of benign and malicious Android apps for our experiments with the help from our research partner Leopard Mobile Inc. Our experiment results demonstrate that the proposed system has accurate security analysis on contracts. Furthermore, we keep our research results and experiment materials on \textsl{http://R2D2.TWMAN.ORG}.

\end{abstract}

\section{Introduction}
\begin{figure}[hbtp]
	\centering
	\includegraphics[width=.8\columnwidth]{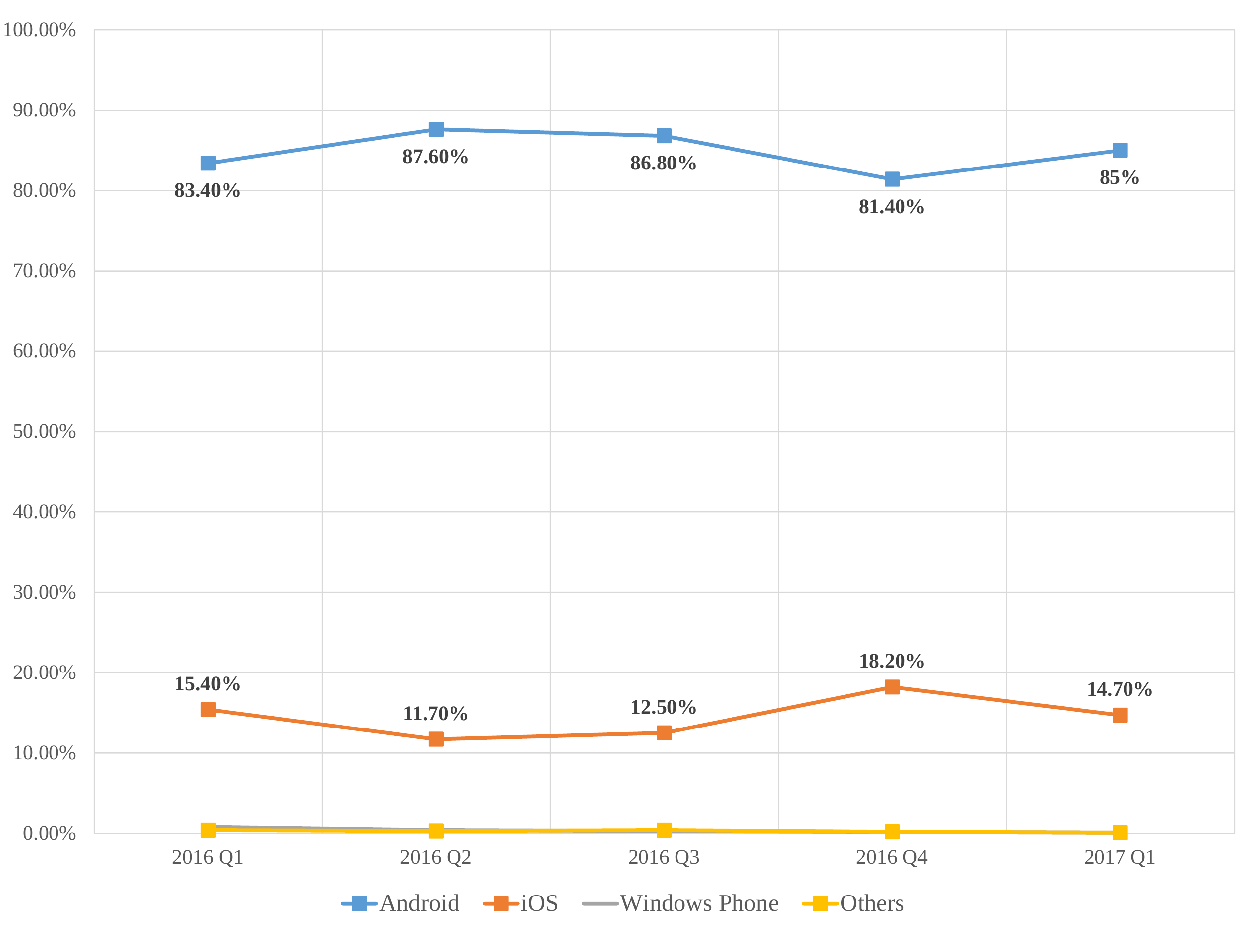}
	\caption{World Smartphones OS Market Share.}\label{fig: F01}
\end{figure}

\begin{figure}[hbtp]
	\centering
	\includegraphics[width=.8\columnwidth]{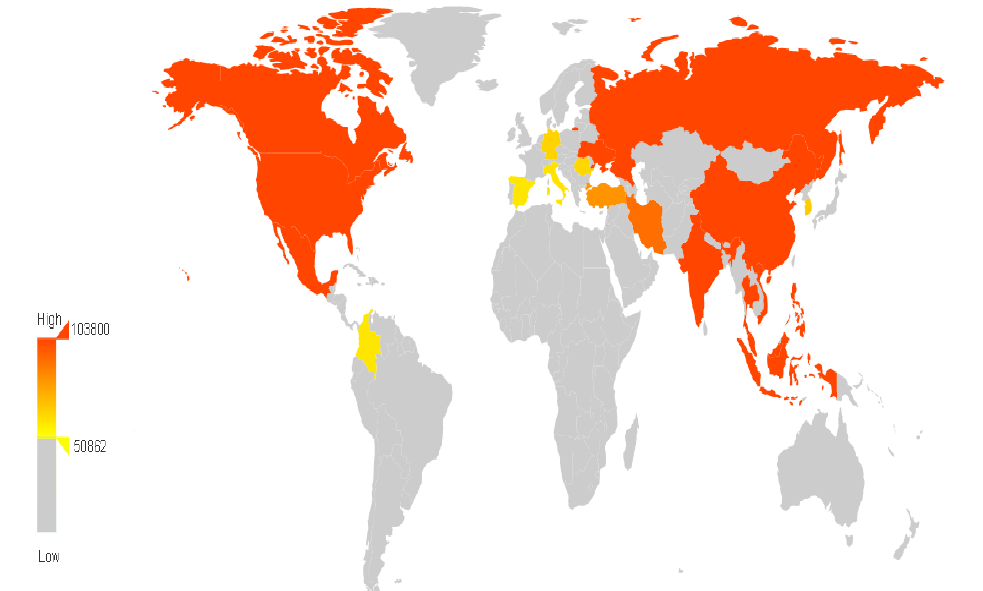}
	\caption{Statistics collected from our back-end system in January 2017 in the US, UK and France.}\label{fig: F02}
\end{figure}

Nowadays, smartphone has become a daily necessity in our life. In the smartphone market, Android is the most commonly used operating system (OS), and it is still expanding its market share. According to the report by International Data Corporation (IDC) in  2018,  the  market  share  of  Android  in smartphone market was approximately 85\% in 2017 Q1 (see Fig. \ref{fig: F01}) \cite{idc2016q2}. Android is featured by its openness; users can choose to download apps from Google Play or third-party marketplace. However due to the popularity and openness, Android has attracted attacker’s attention. Particularly, malicious software (malware) can easily be spread and infects benign Android devices. The Security Report of AV-TEST Institute shows that while the number of malware increased from 17 million in 2005 to over 600 million in 2016, the percentage of Android malware had a significant increase from 3.19\% in 2015 to 7.48\% in 2016 Q2. Among them, Trojans targeting at stealing user data occupied 97.49\%. We also found that Android malware has dominated the market with 99.87\% on the number of malware on smartphone platform \cite{avtest-1516}. Fig. \ref{fig: F02} shows the statistics collected from our back-end system in January 2017. In countries such as the United States, United Kingdom, and France, etc., more than 50,000 users were infected daily. Moreover, according to our mobile security report \cite{cmcm-2017}, it shows that the number of Android malware increased sharply from 1 million in 2012 to 2.8 million in 2014. In 2015, the number of Android malware detected was three times than the one in 2014 (over 9.5 million). In 2016 the number of Android malware achieved more than 17 million. In the first half of 2017, there was more than 10 million Android malware found in \textsl{Security Master} and \textsl{Clean Master}. It has been five years that the number of malware found in Android exceeded the number of malware found in Windows OS over the past 21 years \cite{trendmicro16}. It is obvious that we cannot neglect the security problem from Android nowadays. To deal with the serious security problem caused by Android malware, we proposed the colo\textsl{R}-inspired convolutional neu\textsl{R}al networks (CNN)-based Androi\textsl{D} malware \textsl{D}etection, \textsl{R2-D2}, to detect Android malware. \textsl{R2-D2} is different from existing solutions. The \textsl{R2-D2} detection is featured by its end-to-end learning process. More specifically, in contrast to the prior solutions that require manual process of feature selection and parameter configuration, \textsl{R2-D2} can effectively decrease the resource inputs of manpower and computing. With \textsl{R2-D2}, we can now process and analyze tons of real-time data faster than before. Meanwhile, we can also detect unknown Android malware in a more effective way.

\section{Related Work}

\subsection{The Background of Android Malware Analysis}

Android Package (APK) is the format used by Android OS for the distribution and installation of mobile apps. Essentially, APK file is a type of zip-formatted archive file. The structure of APK can be dissected as follows:
\begin{itemize}
\item META-INIF/: containing the information description from Java jar file. 
\item res/: containing the resource document. 
\item libs/: containing the .so library from Android Native Development (NDK). 
\item AndroidManifest.xml: contains the configuration file about the authorization and service. 
\item classes.dex: containing the dalvik byte-code, i.e., Android execution file. 
\item resource.asc: containing the binary resource file after compilation. 
\end{itemize}

\begin{figure}
	\centering
	\includegraphics[width=.89\columnwidth]{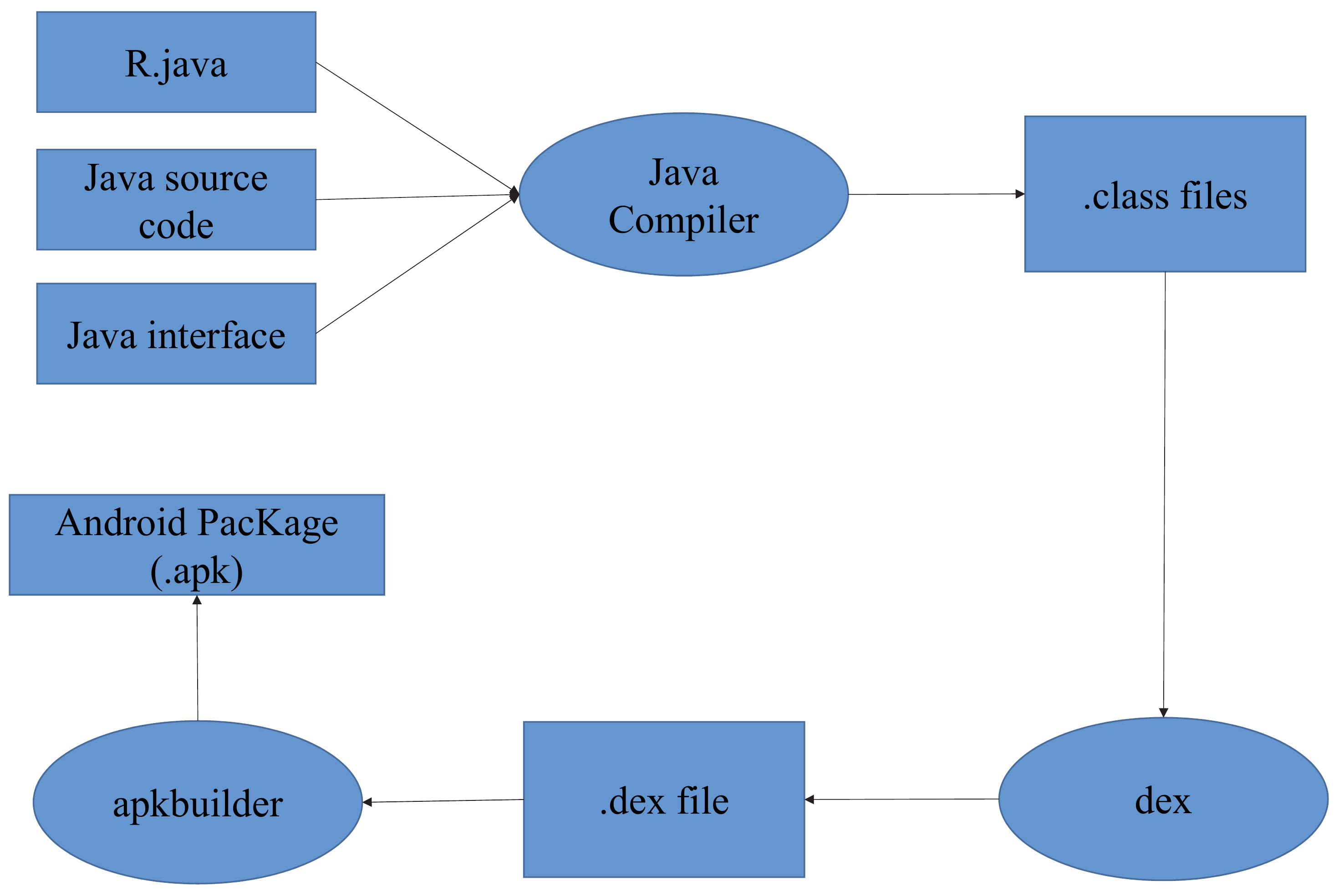}
	\caption{The compilation process of Android APK.}\label{fig: F03}
\end{figure}

Fig. \ref{fig: F03} shows the compilation process of Android APK. First we compile all of the Java source code (including R.Java and Java Interface etc.), generating the .class files. After that, all of the .class files are transformed into dex format generating .dex files supported by Davik VM (Dalvik VM is alike Java VM). It is a bytecode compiler provided on Android phones, mainly developed by company such as Google etc., with the minimal requirement on the resource. In other word, Dalvik VM can be run on the mobile devices with limited computing and memory resource but with acceptable performance. The .dex file is a Java application program, which is in Dalvik executable format. It is all packaged as Android Package (.apk) file by apkbuilder.

One of the most common approaches for Android malware detection is static analysis (signature-based) method. It works through reverse-engineering tools such as \textsl{Apktool}\footnote{https://ibotpeaches.github.io/Apktool/install/}, \textsl{backsmali}\footnote{https://bitbucket.org/JesusFreke/smali/downloads/}, \textsl{dex2jar}\footnote{https://github.com/pxb1988/dex2jar}, and \textsl{JD-GUI}\footnote{https://github.com/java-decompiler/jd-gui}, to decompile Android App to access the source code from classes.dex. Furthermore, \textsl{AXMLPrinter2}\footnote{https://code.google.com/archive/p/android4me/downloads} is used to analyze the AndroidManifest.xml to access the permission for using Android Apps. Another category of approach for Android malware detection is the dynamic analysis (behavior-based) method. The idea behind this category is to continuously track the system communications and network connections or messages by emulating Android apps through techniques such as sandbox (eg. \textsl{Droidbox}\footnote{https://github.com/pjlantz/droidbox}), virtual machine and other operating environment in order to require the dynamic analysis of the behavior model. The security experts then analyze and inspect the records to see whether the malicious behaviors occur. If so, virus pattern and black/white-list are generated for pattern matching and for capturing the malware. However, anti-analysis techniques such as obfuscation, encryption, and anti-debugging are also proposed by hackers to hide the malicious behaviors to escape from the detection \cite{ccs14}.  In fact, the above approaches are useless when the malware programmers make modification on the initial components in malware, which can evade the detection. 

\begin{figure}
	\centering
	\includegraphics[width=.8\columnwidth]{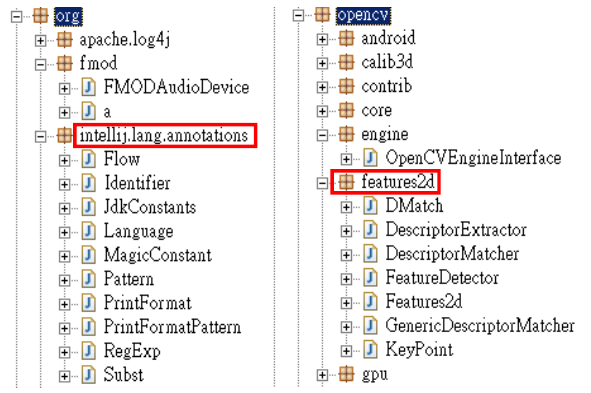}
	\caption{The popular class naming format in benign Android apps.}\label{fig: F04}
\end{figure}
\begin{figure}
	\centering
	\includegraphics[width=.9\columnwidth]{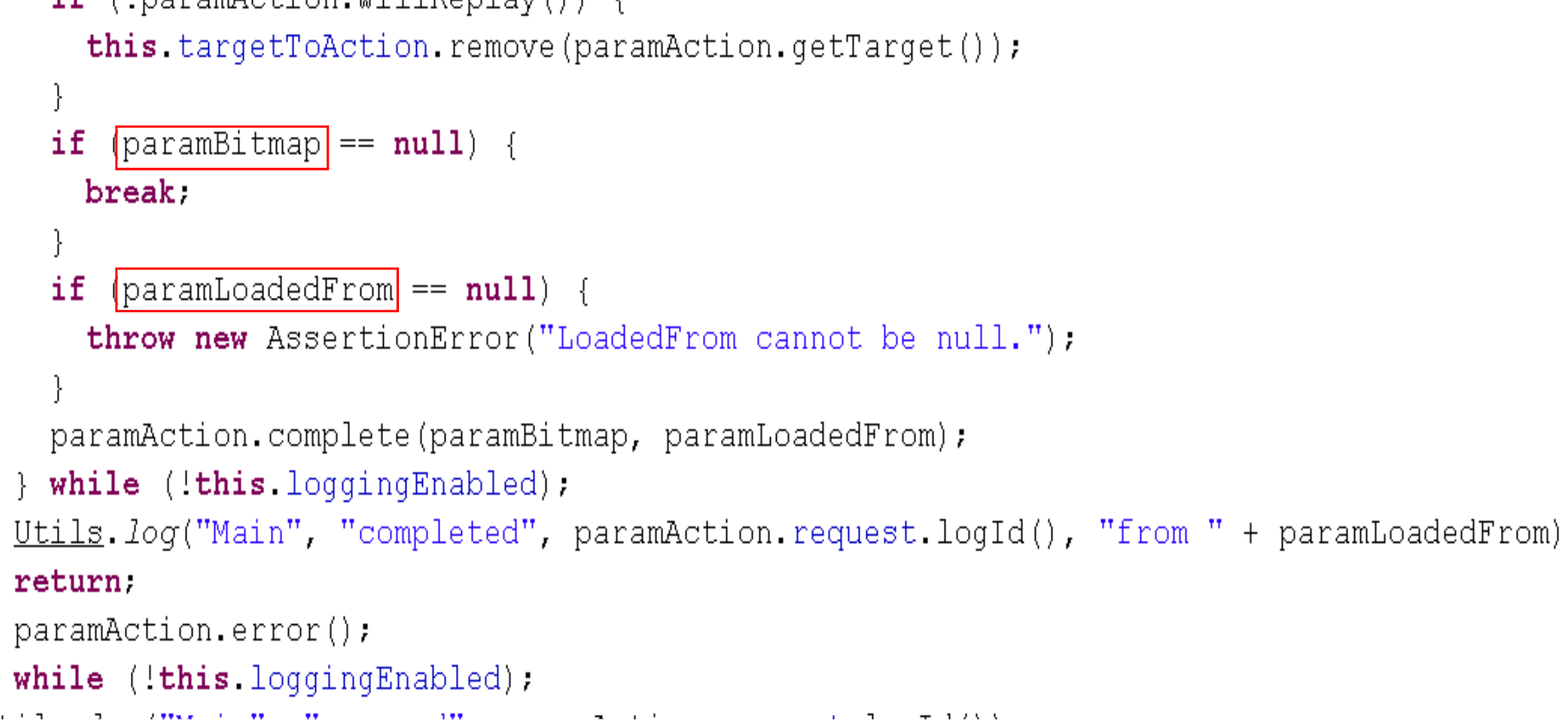}
	\caption{The common way for declaring the data variables.}\label{fig: F05}
\end{figure}
\begin{figure}
	\centering
	\includegraphics[width=.9\columnwidth]{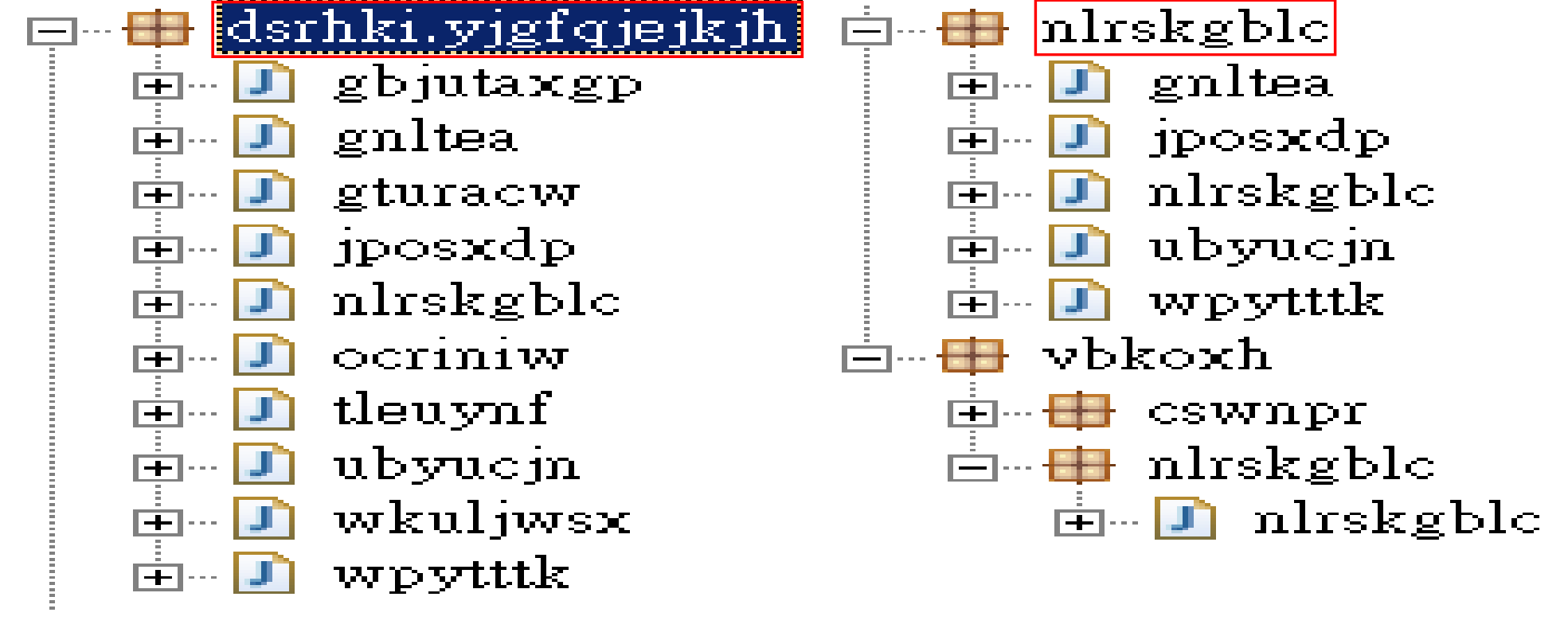}
	\caption{Certain random naming format will be used by opfake family.}\label{fig: F06}
\end{figure}
\begin{figure}[htp]
	\centering
	\includegraphics[width=.9\columnwidth]{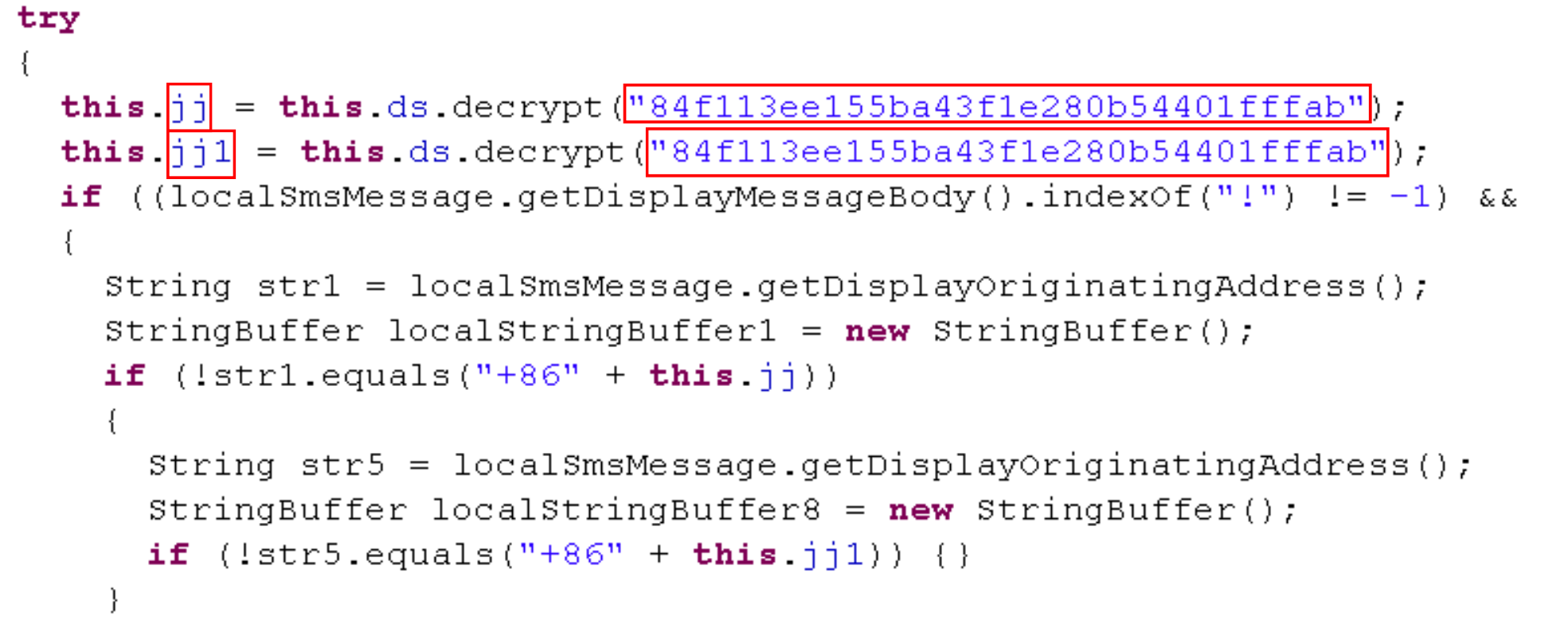}
	\caption{Encrypt variables by the fakeinst family.}\label{fig: F07}
\end{figure}

Fig. \ref{fig: F04} shows the popular class naming in benign Android apps (e.g. intellij.annotations and features2d). Fig. \ref{fig: F05} shows the common way for declaring the data variables (e.g. paramBitmap and paramLoadedForm). However, to evade the detection, certain random naming (e.g. dsrhki.yjgfqjejkjh and nlrskgblc) will be used by “opfake” family (shown in Fig. \ref{fig: F06}). Fig. \ref{fig: F07} shows the “fakeinst” family that can automatically encrypt variables (e.g. jj = 84f113ee155ba4f1e280b54401fffab and jj1 = 84f113ee155ba43f1e280b54401fffab).

\subsection{Machine Learning-based Malware Detection}
Machine learning is a technique that enables the machine to learn the pattern, build the model, and make the prediction by seeing only the raw data. Currently, the most widespread malware attacks are still "\textsl{exploit attack}" and "\textsl{privileged escalation}". As a result, most of the machine learning-based malware detection still use the features such as the Android apps permissions, API invocation and control flow graphs (CFG) to distinguish between the benign and malicious apps. Usually, SVM and random forest are used to build the model to distinguish between benign and malicious apps. 

For example, DroidMiner \cite{DroidMiner} has a two-step flow-chart to represent the behaviours of Android apps and capture the execution logs behind the apps. After that, rectors are clustered to identify the similar code snippet. Lei Chen et al. \cite{TDSC15} extracted features from Android API invocation as a reverse engineering approach. They apply normalization procedure to the extracted featuring and perform logistic regression on the extracted feature to detect Android malware. Moreover, the dataset used by the current research-oriented machine learning-based Android malware detection is rather small. Most of existing solutions train their detection models sorely based on small datasets. This cause the practicality problem because the real world malware may have distinguished distinctive behaviors and the existing detection models usually cannot successfully identify the Zero-day exploit malware by correlating the API invocation and Android permissions. 

\subsection{Deep Learning}
Recent success in deep learning research and development attracts people’s attention \cite{DeepMind16}. In 2015, Google released Tensorflow \cite{tensorflow16}, a framework of realizing deep learning algorithm. Deep learning is a specific type of machine learning. More specifically, deep learning is an artificial neural network, in which multiple layers of neurons are interconnected with different weights and activation functions to learn the hidden relationship between input and output. Intuitively, input data is fed to the first layer that generates different combinations of the input \cite{LeCun15}. After the activation function, these combinations are fed to the second layer, and so on. Under the above procedures, different combinations of the outputs from previous layer can be seen as different representation of features. The weights on links between layers are adjusted according to backward propagations, depending on the distance or less function between true output label and the label calculated by neural network. Note that deep learning can be seen as a neural network with a large number of layers. After the above learning process via multiple layers, we can derive a better understanding and representation of distinguishable features, enhancing the detection accuracy \cite{Goodfellow16}.

In addition to deep neural networks, the most well-known deep networks are convolutional neural networks (CNN). The representation of CNN includes AlexNet, VGG, GoogleNet, Inception-v3 and ResNet \cite{Alexnet12}\cite{VGG15}\cite{Inception16}\cite{ResNet16}. More specifically, CNN is composed of hidden layers, fully connected layers, convolution layers, and pooling layers. The hidden layers are used to increase the complexity of the model. If the same number of neural is associated with the input image, the number of parameters can be significantly reduced, adapting to the function structure much properly.

\subsection{Deep Learning-based Malware Detection}
\cite{DL4MD16} proposes a malware detection in which the Windows API inquiry generates a corresponding ID, which is treated as the input of the deep-learning architecture (eg. stack of Auto-Encoders), and then it fine-tunes the model parameters. \cite{MALWARE16} is a method that it works on feature extracting first, such as contextual byte features, PE import features, string 2d histogram features, and PE metadata features. Then, the extracted features are fed to the deep neural network (DNN). With the training of two hidden layers, it is categorized. \cite{SIGCOMM16} uses static analysis to extract features such as required permission, sensitive API, and also uses dynamic analysis to extract features such as "action dex class load", "action recrent" and "action service start", from 500 samples for about 200 features as the input for the deep belief network (DBN). 

Noted, there is similarity between the execution logic of Android malicious apps and the order of functions being called. Thus, in addition to the aforementioned solutions that apply DNN to malware analysis based on "exploit attack" and "privileged escalation", another category of malware detection relies on n-gram analysis on byte-code or op-code. For example, \cite{NGramBytecode1} and \cite{NGramBytecode2} first calculate the n-grams on the binary byte-code and then perform the malware detection based on k-nearest neighbor. \cite{NGramopcode} proposes to do reverse-engineering first and then analyzes op-code. In addition, one more category of the malware detection relies on transforming malware into the images. For example, \cite{Malwareimagess} proposes to first transform binary byte-code into gray scale image and then applies pattern recognition to the gray scale image. 

All of the above methods have achieved a certain level of detection accuracy. Deep learning once is seemed as the cure for the above problem. However, a pre-processing step, such as feature engineering, is still needed before the model is learnt. Also, the dataset used by the current research-oriented machine learning-based Android malware detection is rather small. The dataset for training the model usually cannot reflect real-world malware accurately. On the other hand as mentioned in the introduction, the number of malware increased dramatically. Even worse, more and more anti-debugging techniques are discovered. The size of dataset used for training the model also has significant impacts on the detection accuracy and the computing efficiency in the training process. Here, we particularly note that despite the detection accuracy of the n-gram approach, N-gram approach consumes substantial computing resources and time for handling the dynamic growth of the model parameters required, implying the impracticality \cite{YLeCun15}. 

However, if we have limited computing resources and time, CNN is able to handle the explosive data growth because the increasing number of parameters does not imply the growth of computing resource and time required. Recently, \cite{DASPS17} also proposes deep learning-based malware detection, where the sequences of the op-code are encoded as one-hot vectors for the input of CNN. However, this method needs to dissemble the Android apps via reverse-engineering tools (backsmali) for deriving smali source code from classes.dex, and therefore cannot handle malware with encryption (as shown in Fig. \ref{fig: F06} and Fig. \ref{fig: F07}) or obfuscation \cite{rsa15}, \cite{bhusa12}, \cite{defcon22}.

\section{OUR PROPOSED MECHANISM: R2-D2}
\subsection{The Characteristic of Our Methodology}
According to the second chapter and our previous research projects, the mainstream is still using the static analysis to identify features via reverse engineering the source code of Android apps and using dynamic analysis to collect the order of API calls. Though a certain degree of detection accuracy is presented, the reality is that due to the rapid iteration, and various reverse engineering analytic method, we found that the amount of unknown sample (called gray sample) still cannot be detected by current methods. It is difficult to improve the detection efficiency and its accuracy \cite{cmcm-2017}\cite{hitcon}.
If unknown new Android Apps are scanned on user’s Android devices, no matter it is benign or malware, we may need to upload the APK to our back-end and then undergo manual or automatic feature extraction. After that, we can start identification. This process consumes a large amount of network traffic on user side, and it is not real-time enough on efficiency in the process. If gray sample can be detected fast enough, we can decrease the tendency for Android device to be attacked.

\begin{figure}[htp]
	\centering
	\includegraphics[width=1\columnwidth]{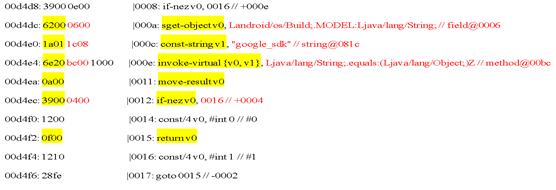}
	\caption{Dalvik bytecode example code which detect the Android device’s product model.}\label{fig: FDalvik1}
\end{figure}

To solve the problem from traditional time-consuming and resource-consuming method, according to our research of \cite{NGramBytecode1}, \cite{NGramBytecode2} and \cite{NGramopcode}, we found that if Android apps execute similar behavior, its API will perform similar order of opcode; for example, as shown in Fig. \ref{fig: FDalvik1}, "\textsl{6200}" represents the sget-object function "\textsl{reads the object reference field identified by the field\_id into vx}", and "\textsl{0600}" represents the "\textsl{field address of android/os/Build;.MODEL}". "\textsl{6e20}" represents the invoke-virtual function: "\textsl{invokes a virtual method with parameters}", for extracting the value from v0 and v1 to compare, where v1 is from previous line const-string v1 of "\textsl{1a01}" to get the value of google sdk ("\textsl{puts reference to a string constant identified by string\_id into vx}"). 

\begin{figure}[htp]
	\centering
	\includegraphics[width=.95\columnwidth]{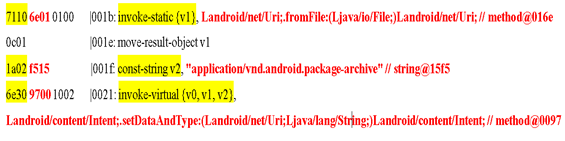}
	\caption{Dalvik bytecode example code which try to install other apk.}\label{fig: FDalvik2}
\end{figure}

Fig. \ref{fig: FDalvik2} gives another example of what the Dalvik bytecode shows if the Android applications try to install APK package by itself. This is done by using "\textsl{invoke-static {v1}}" (7110) as the Android API method \textsl{"Landroid/net/Uri;.parse}" to create an URI (6e01), which parses the given encoded URI string, and moves reference {v1} and "\textsl{move-result-object v1}" (0c01) to the string const-string v2 (1a02) specified by the given index into the specified register {v2} as"\textsl{application/vnd.android.package-archive}" (f515), setting the data with an explicit MIME data type. We will further explain the hexadecimal from byte-code to rgb color code through the rule in section III.B.

Based on the studies of image detection in CNN \cite{Alexnet12}\cite{VGG15}\cite{Inception16}\cite{ResNet16} and the method proposed by \cite{Malwareimagess}, we aim to figure out the hidden relationship between the program execution logic and the order of function calls behind the malware by taking advantage of the CNN in order to accurately detect known and even unknown malware. Since the bytecode in classes.dex and rgb color code are both hexadecimal, we proposed an algorithm that can convert the bytecode as rgb color code into color image (called \textsl{Android color images}). Then we applied CNN to the \textsl{Android color images} for detecting the Android malware. Our proposed system, \textsl{R2-D2} (colo\textsl{R}-inspired convolutional neu\textsl{R}al network-based androi\textsl{D} malware \textsl{D}etection), works by decompressing Android apps (APK) and then translating the dalvik executable classes.dex into \textsl{Android color images}.

Based solely on Android color images, our \textsl{R2-D2} system achieves the end-to-end training and is able to make accurate detection. The reason we are eager to achieve this end-to-end training is that, we realize a huge amount of human labor will be spent on feature engineering and detection modeling. To ease the model training, we adopt the above deep learning approach to construct an end-to-end learning-based Android malware detection. Our \textsl{R2-D2} possesses the following advantages \cite{owasp}:

\begin{itemize}
\item \textsl{R2-D2} translates classes.dex into RGB color images, without modifying the original Android apps and without extracting features from the apps manually in advance. It can complete a translation within 0.4 second. Such translation is also featured by the fact that more complex information in the Android apps can be preserved in the color image with 16777216 colors (with 24 bit pixels per sample) compared to the gray scale image with only 256 colors (with 8 bit pixels per sample).

\item With the fully connected network infrastructure of DNN, it can deal with rapid iterated malware with its large amount of parameters, but the local receptive fields and shared weights of CNN make it better for more complex structure. It not only decreases the amount of parameters, but also reflects the complexity of Android malware. Comparing to current method, it saves a lot of time without huge computation.
\item The auto-image feature extraction in CNN does not extract features directly from image for pattern recognition. Instead, the raw pixels are represented by multi-dimensional matrices. Then through the calculation of filter and the size of stride in convolution layer, and non-linear activation functions in pooling layer, it can enhance the permutation relation between data.
\item We only need the classes.dex in Android App, so if there is unknown new App appears, we only need to transform it to Android color image. The image size is about 10-50kb. Compared to uploading the App to the back-end and then processing the extraction and identification, \textsl{R2-D2} can not only save the traffic loading on user side, but also reduce the resources and speed up the processing efficiency. We will further explain the system architecture and its process in section III.C and Fig. \ref{fig: F14}.
\end{itemize}

\subsection{The Core Technology of Our Methodology}

\begin{figure}[hbtp]
	\centering
	\includegraphics[width=.85\columnwidth]{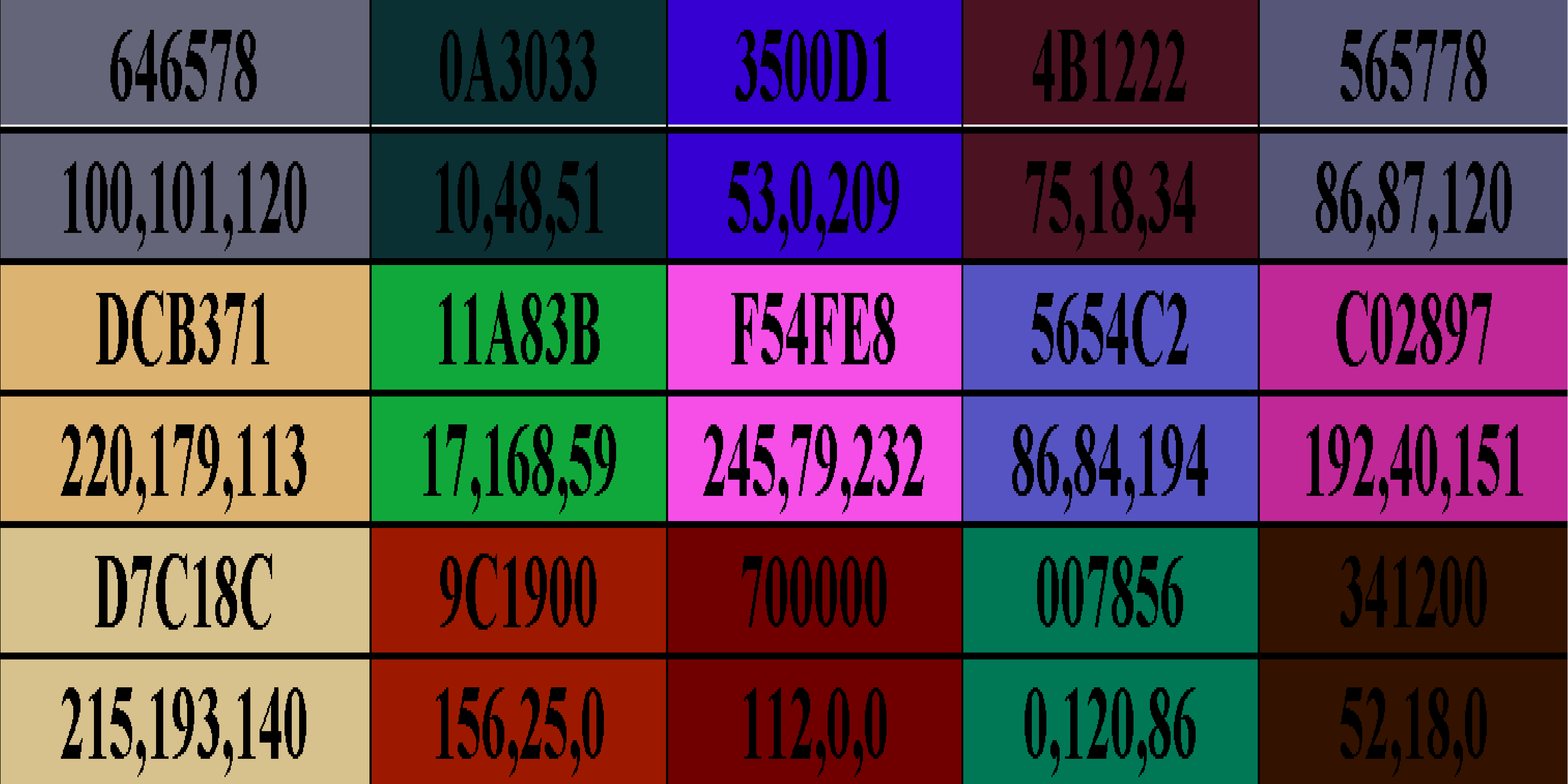}
	\caption{The result of present byte-code as rgb color code.}\label{fig: F09}
\end{figure}

\begin{figure}[hbtp]
	\centering
	\includegraphics[width=.77\columnwidth]{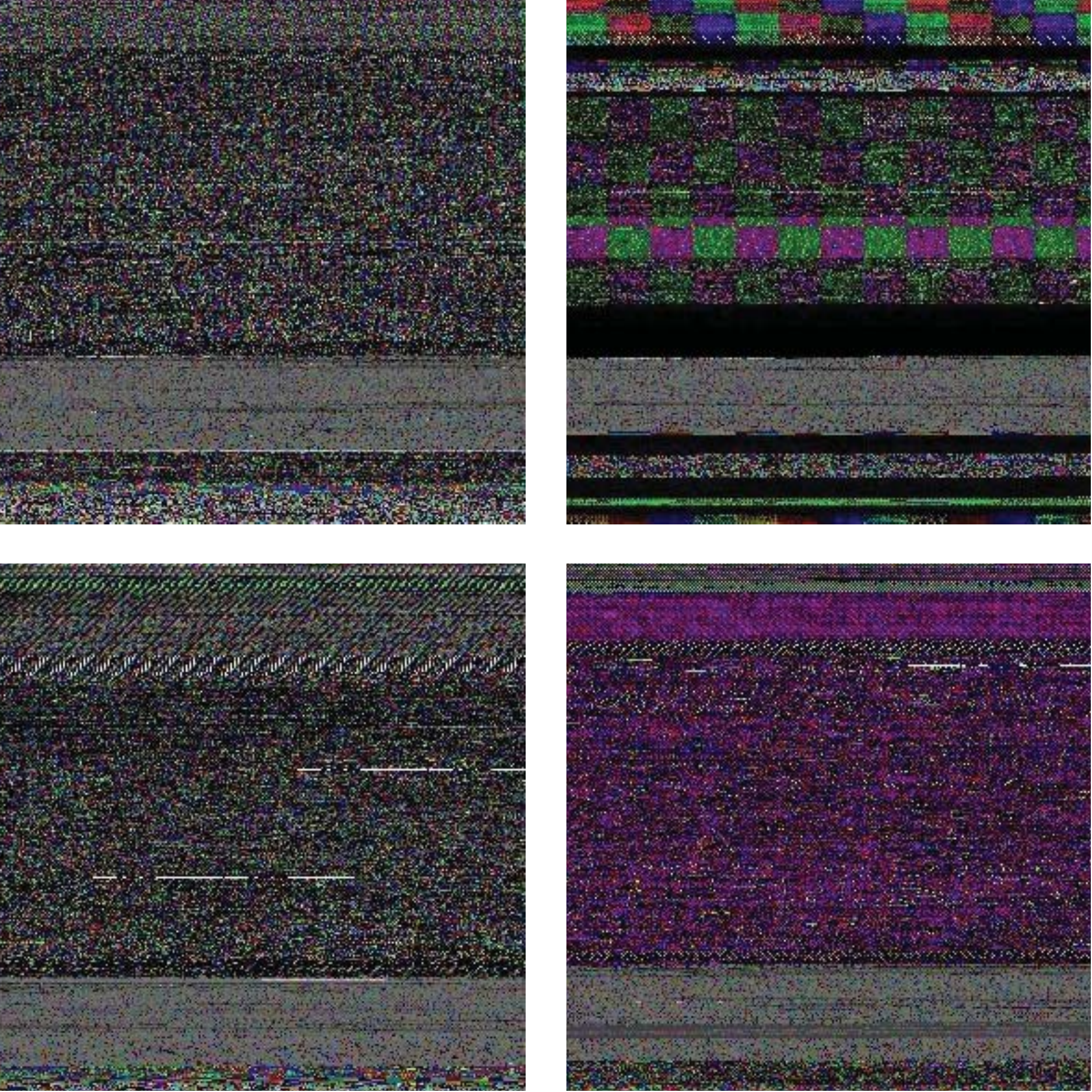}
	\caption{The color images example of benign Android apps.}\label{fig: F11}
\end{figure}

\begin{figure}[hbtp]
	\centering
	\includegraphics[width=.77\columnwidth]{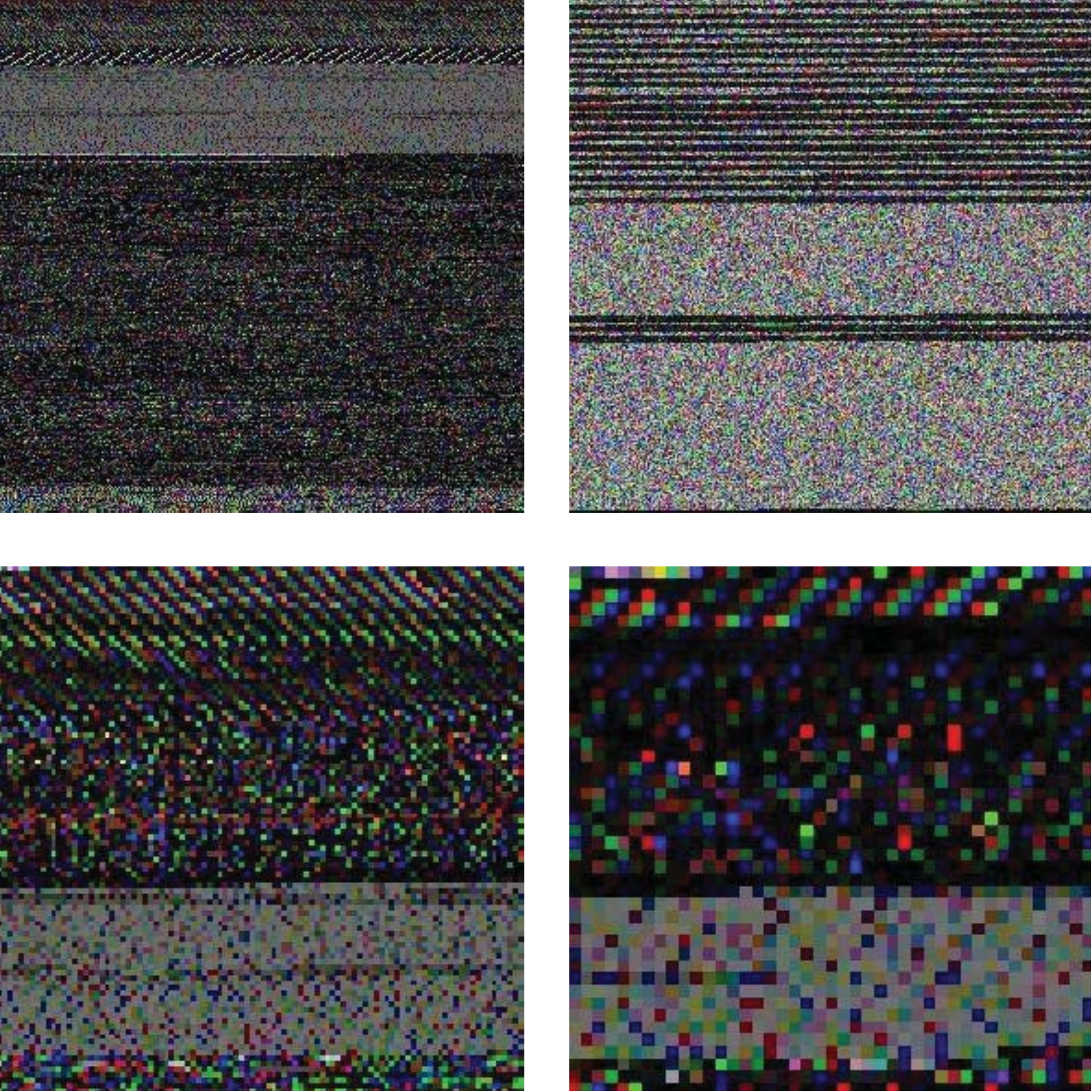}
	\caption{The color images example of malcious Android apps.}\label{fig: F12}
\end{figure}

In the followings, we explain the algorithm procedures of \textsl{R2-D2} in more details. First of all, we collected the Android apps from our research partner's (\textsl{Leopard Mobile Inc.}) original back-end detection system, the Android apps were classified as benign and Trojan, RiskTool, HackTool, AdWare, Banker, Clicker, Downloader, Dropper, FakeAV, Monitor, SMS, Spy, Ransom, Exploit, and BackDoor. Then, we decompressed the apps to retrieve the classes.dex and presented as byte-code. We then mapped the hexadecimal from byte-code to rgb color code through the rule (e.g., 646578 = (R:100, G:101, B:120), 0A3033 = (R:10, G:48, B:51), 3500D1 = (R:53, G:0, B:209) and 4B1222 = (R:75, G:18, B:34) and 565778 = (R:86, G:87, B:120) etc.) shown in Fig. \ref{fig: F09}. Finally, we reached an Android color image (shown in Fig. \ref{fig: F11} and \ref{fig: F12}, which were four different Android benign and malware apps), and the images are fed to CNN and training a model to detection Android malware. 


\begin{figure}[htp]
	\centering
	\includegraphics[width=.8\columnwidth]{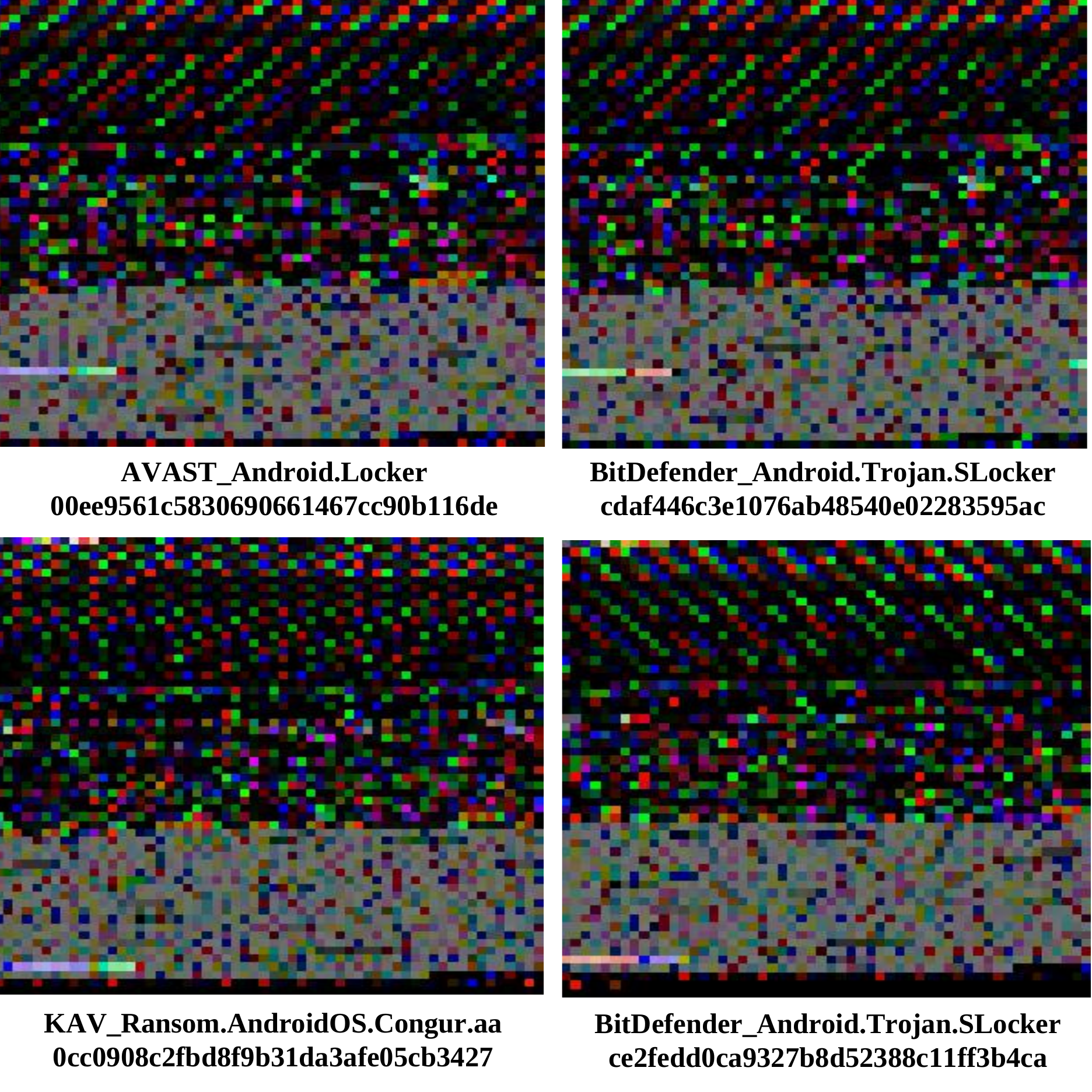}
	\caption{The color images example of Android malware family apps.}\label{fig: F28}
\end{figure}

Meanwhile, we did an image distance test on our Android malware image. The result shows that samples of the same Android malware family share similarity in their visual patterns that are close to each other in the sense of distance from Levenshtein, RMS deviation and MSE image distance. The results validated our  proposed  \textsl{Android  color  image} algorithm that is suitable for the classification with CNN (e.g., 00ee9561c5830690661467cc90b116de of Android:Jisut-JY [Trj] from AVAST and cdaf446c3e1076ab48540e02283595ac of Android.Trojan.SLocker.IS from BitDefender are 54.59\% similar, 0cc0908c2fbd8f9b31da3afe05cb3427 of Trojan-Ransom.AndroidOS.Congur.aa from Kaspersky and ce2fedd0ca9327b8d52388c11ff3b4ca of Android.Trojan.SLocker.IS from BitDefender are 54.94\% similar etc.), as shown in Fig. \ref{fig: F28}. Although the fine-grained sense is not accurate, Android malware image visual mode of this similarity can help us quickly classify Android malware, while reducing labor costs \cite{ruxcon}.

\begin{figure}[hbtp]
	\centering
	\includegraphics[width=.98\columnwidth]{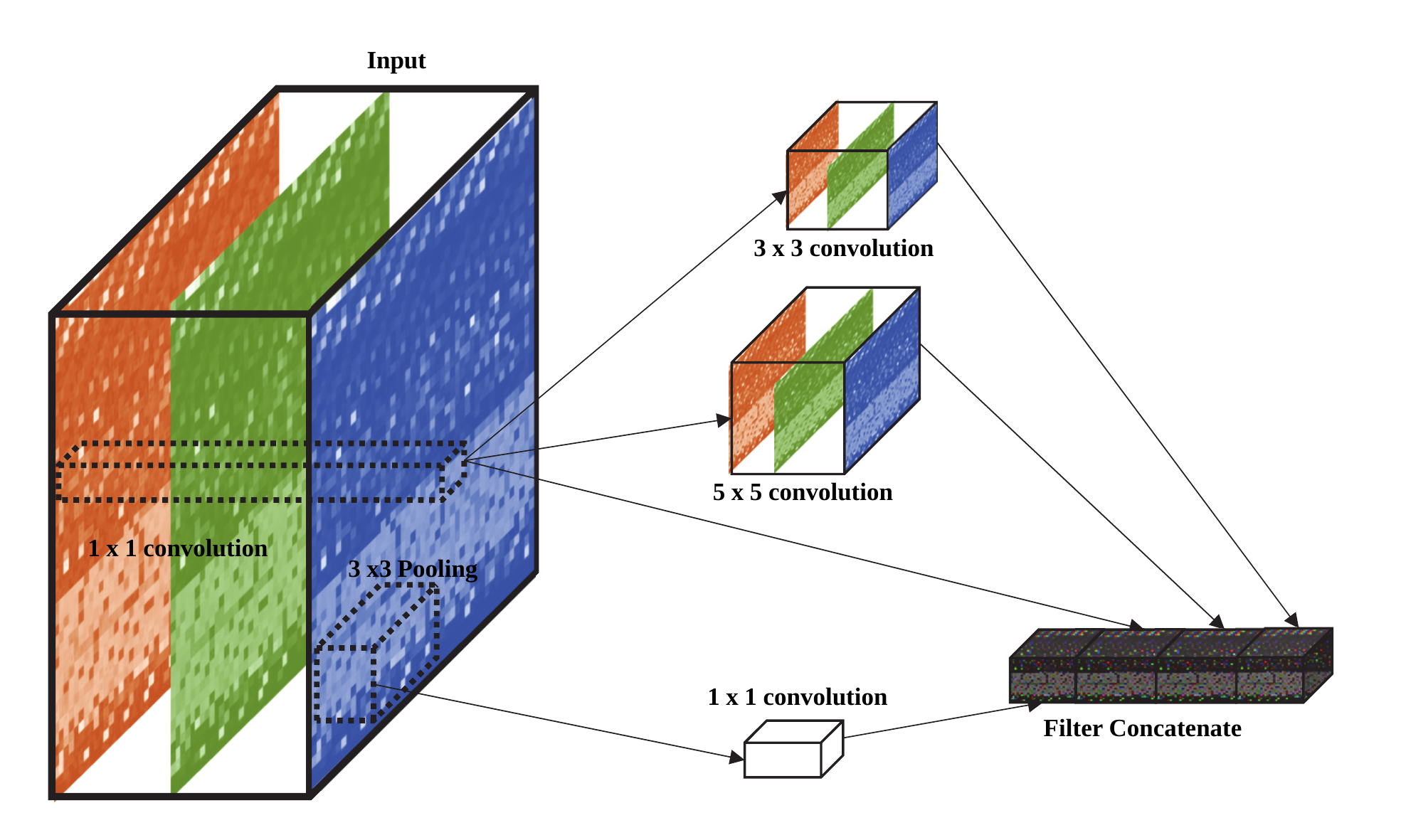}
	\caption{The Inception-v3 structure.}\label{fig: FXX}
\end{figure}

However, we also found that two approaches might be used to escape our Android malware detection.
\begin{itemize}
\item Since the traditional filter size of CNN is 3*3 or 5*5, the uncorrelated bytecode might become correlated when we transform classes.dex into images. The malware may evade the detection by taking advantage of such a mismatch. 
\item Pooling is a common approach in CNN model to reduce the computation overhead significantly in traditional image recognition. The detection engine in our original research inherently uses pooling to achieve the speedup. However, Android color images are not natural images; instead, they are formed from Android source code. Thus, the pooling inevitably destroys the contexts and semantics of the malware code, causing the detection inaccuracy.
\item Noted that CNN trained model is not suited to be embedded in Android App for malware detection due to the reason that the file is too large.
\end{itemize}

To address the above two issues, we did many experiments with CNN models (includes AlexNet, VGG, GoogleNet, and Inception-v3). We found the characteristics of 1x1 convolution in Inception-V3 (shown in Fig. \ref{fig: FXX}). It replaces few filters with a smaller perceptron layer with mixture of 1x1 and 3x3 convolutions, and 1x1 convolutions are specially used before 3x3 and 5x5 convolution to reduce the dimensions. In this way, we can add more non-linearity by having ReLU immediately after every 1x1 convolution and reduce the dimensions inside this “inception module”. Based on \cite{NIN}, 1x1 convolution is equivalent to cross-channel parametric pooling layer, and this cascaded cross channel parametric pooling structure allows complex and learnable interactions of cross channel information. Cross channel information learning (cascaded 1x1 convolution) is biologically inspired because human visual cortex has receptive fields (kernels) tuned to different orientation.

\subsection{The Architecture of Our Methodology}

\begin{figure}[hbtp]
	\centering
	\includegraphics[width=1\columnwidth]{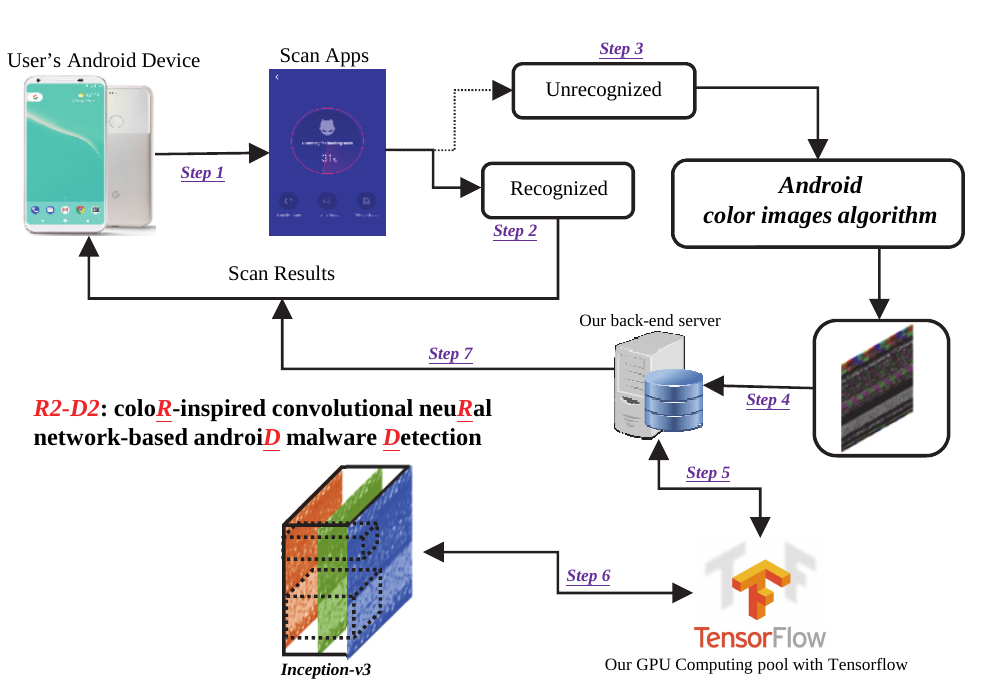}
	\caption{The architecture of R2-D2.}\label{fig: F14}
\end{figure}

Fig. \ref{fig: F14} shows our system architecture. Step 3 to step 6 was mentioned in III.A. Users do not need to upload to consume network traffic. It can improve the efficiency of processing and reduce the computing resources. It can also get rid of the drawbacks of CNN mentioned in III.B while the training model is too large to run on user side.
\begin{itemize}
\item Step 1. User scans the Apps on the Anndroid device.
\item Step 2. If the Apps are all identified. The scanned results will be provided to the users directly.
\item Step 3. If there is unknown App, the system will transform the classes,dex into Android color image.
\item Step 4. Then upload the Android color image to the back-end.
\item Step 5. Feed the Android color image to the GPU computing pool of tensorflow.
\item Step 6. Identify the Apps with the trained Inception-v3 model.
\item Step 7. Send the results to user’s phone.
\end{itemize}

\section{EXPERIMENTAL RESULTS}

\subsection{Experiment Environment and Datasets}

We run it on a 64-bit Ubuntu 14.04, and hardware setting are 128 GB DDR4 2400 RAM and Intel(R) Xeon(R) E5-2620 v4 CPU, NVIDIA TITAN V, TITAN XP and GTX 1080 GPUs; more specifically, the software setting is  the nvidia-docker tensorflow:18.04-py2 on nVIDIA cloud. The research results and the data from our research partner, Leopard Mobile Inc., can be found on the website \textsl{http://R2D2.TWMAN.ORG} (72 GB) \cite{owasp}. The data was collected from Jan. 2017 to Aug 2017. Approximately 2 million of benign and malicious Android apps were collected for our experiments; till now, we have uploaded data and results from Jan. 2017 to May 2017 and continue to train and adjust our model.

\subsection{Evaluations of different deep neural networks optimization}
\begin{figure}[htp]
	\centering
	\includegraphics[width=1\columnwidth]{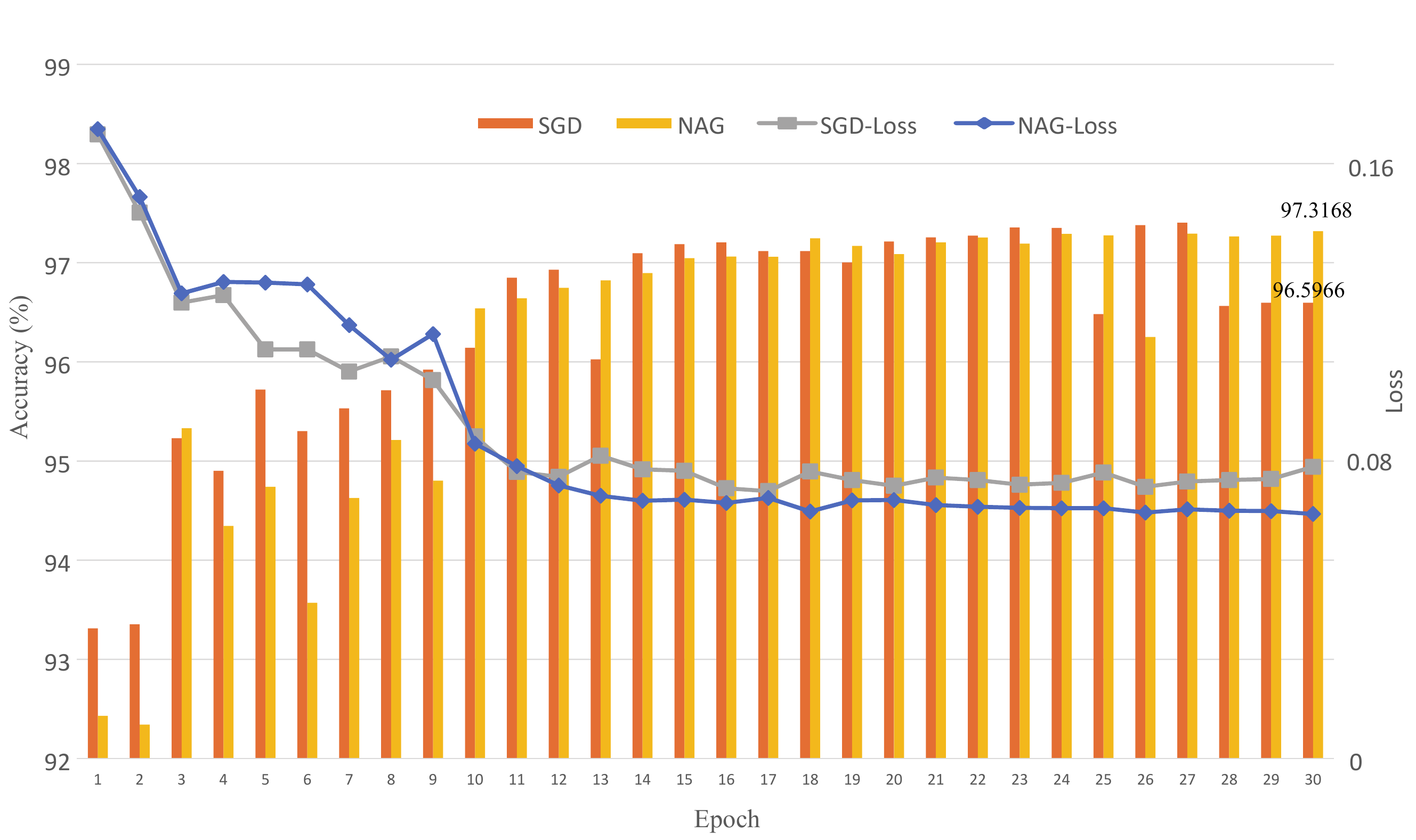}
	\caption{The results of Inception-v3 with the SGD and NAG optimization.}\label{fig: F19}
\end{figure}

\begin{figure}[htp]
	\centering
	\includegraphics[width=1\columnwidth]{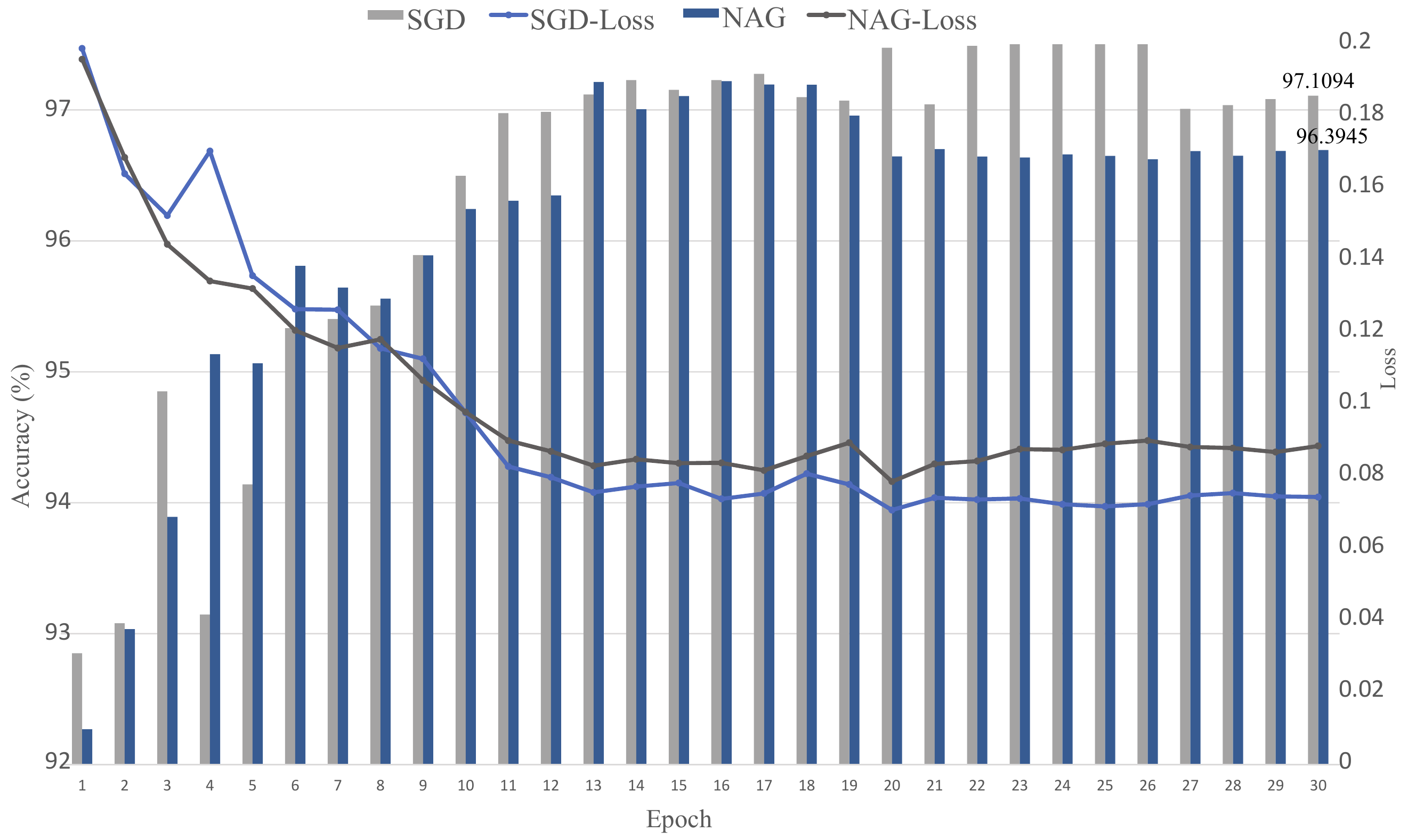}
	\caption{The results of AlexNet with the SGD and NAG optimization.}\label{fig: F20}
\end{figure}

\begin{figure}[htp]
	\centering
	\includegraphics[width=1\columnwidth]{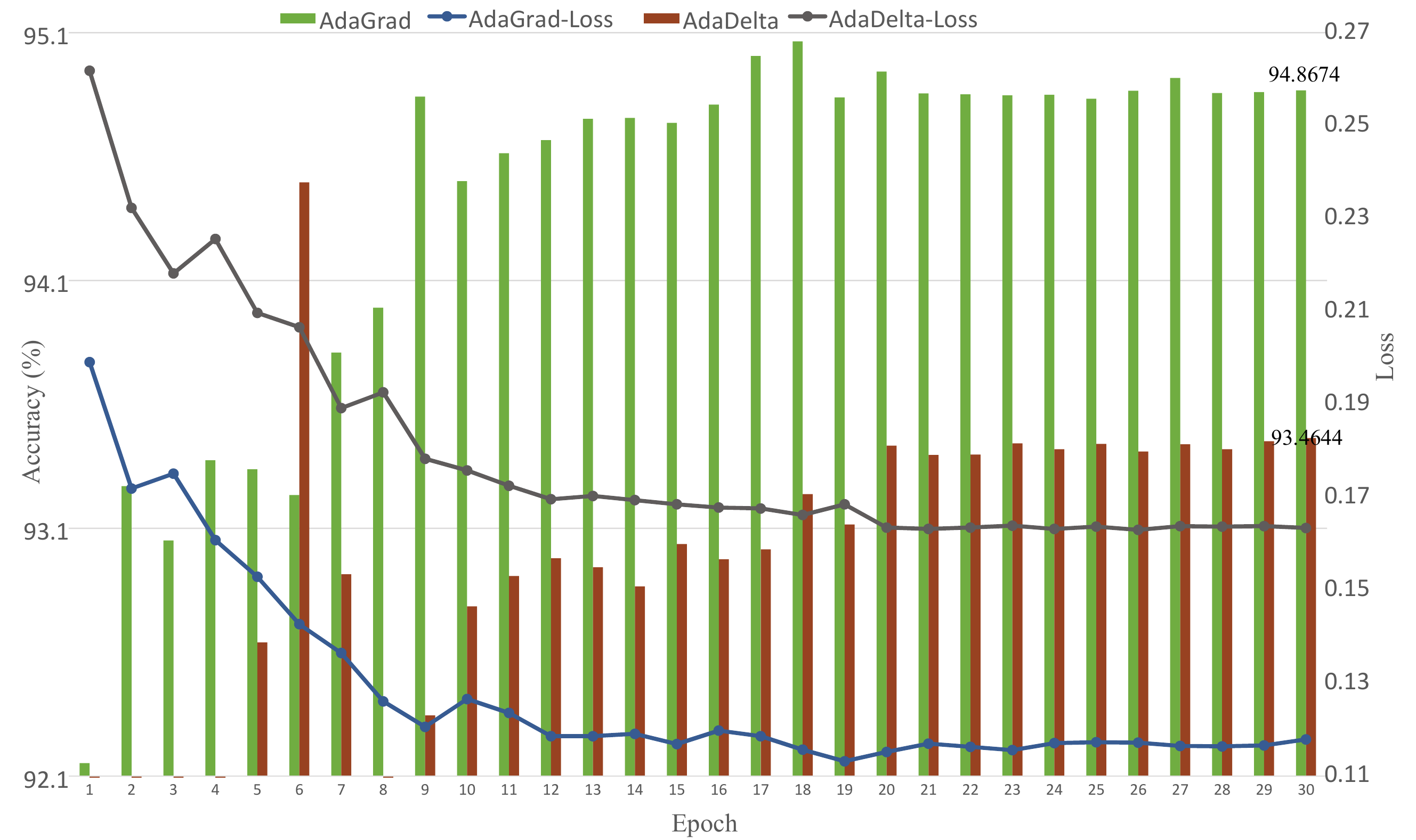}
	\caption{The results of Inception-v3 with the AdaDelta and AdaGrad optimization.}\label{fig: F21}
\end{figure}

\begin{figure}[htp]
	\centering
	\includegraphics[width=1\columnwidth]{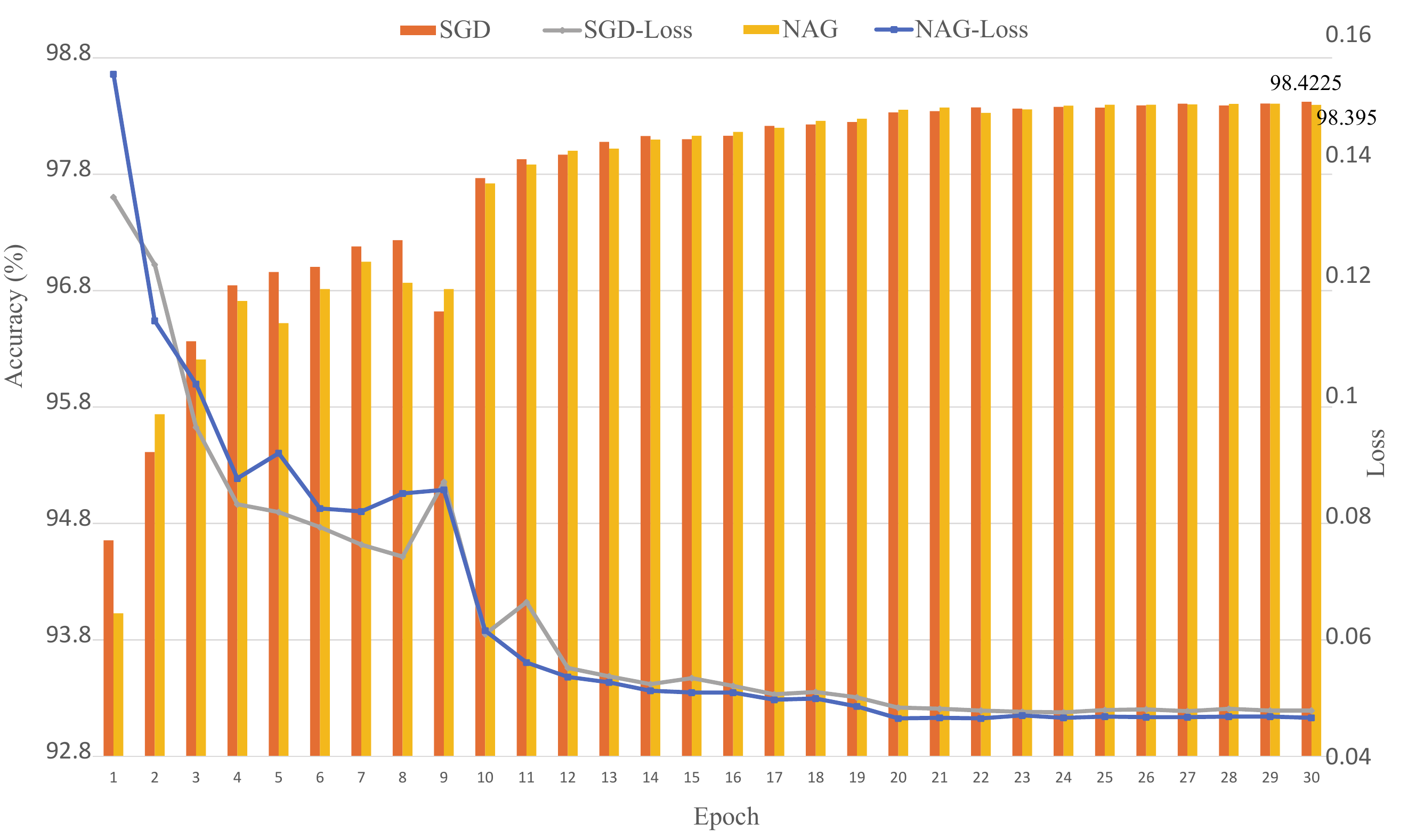}
	\caption{The results of Inception-v3 with the SGD and NAG optimization.}\label{fig: F22}
\end{figure}

Based on our collected data, we evaluate the detection accuracy and performance with different network model (e.g. Alexnet, Googlenet, Inception-v3) and optimization. Noted that the learning rate is fixed to be 0.01, the optimization methods used are stochastic gradient descent (SGD), Nesterov Accelerated Gradient (NAG), AdaDelta and AdaGrad. From our experiment, we found that Inception-v3 (shown in Fig. \ref{fig: F19}) is almost always better than Alexnet (shown in Fig. \ref{fig: F20}) and other models. It also verified that adopting Inception-v3 can solve the drawbacks that mentioned before in our \textsl{R2-D2} method. With such observation, we further fine-tuned and compared Inception-v3 with AdaDelta and Inception-v3 with AdaGrad (see Fig. \ref{fig: F21}), and found that SGD is best suitable for our use (see Fig. \ref{fig: F22}). In particular, it resulted in the sharpest increase in accuracy and sharpest decrease in loss. As a result, we reach 98.4225\% and 97.7081\% accuracy.

\begin{figure}[htp]
	\centering
	\includegraphics[width=1\columnwidth]{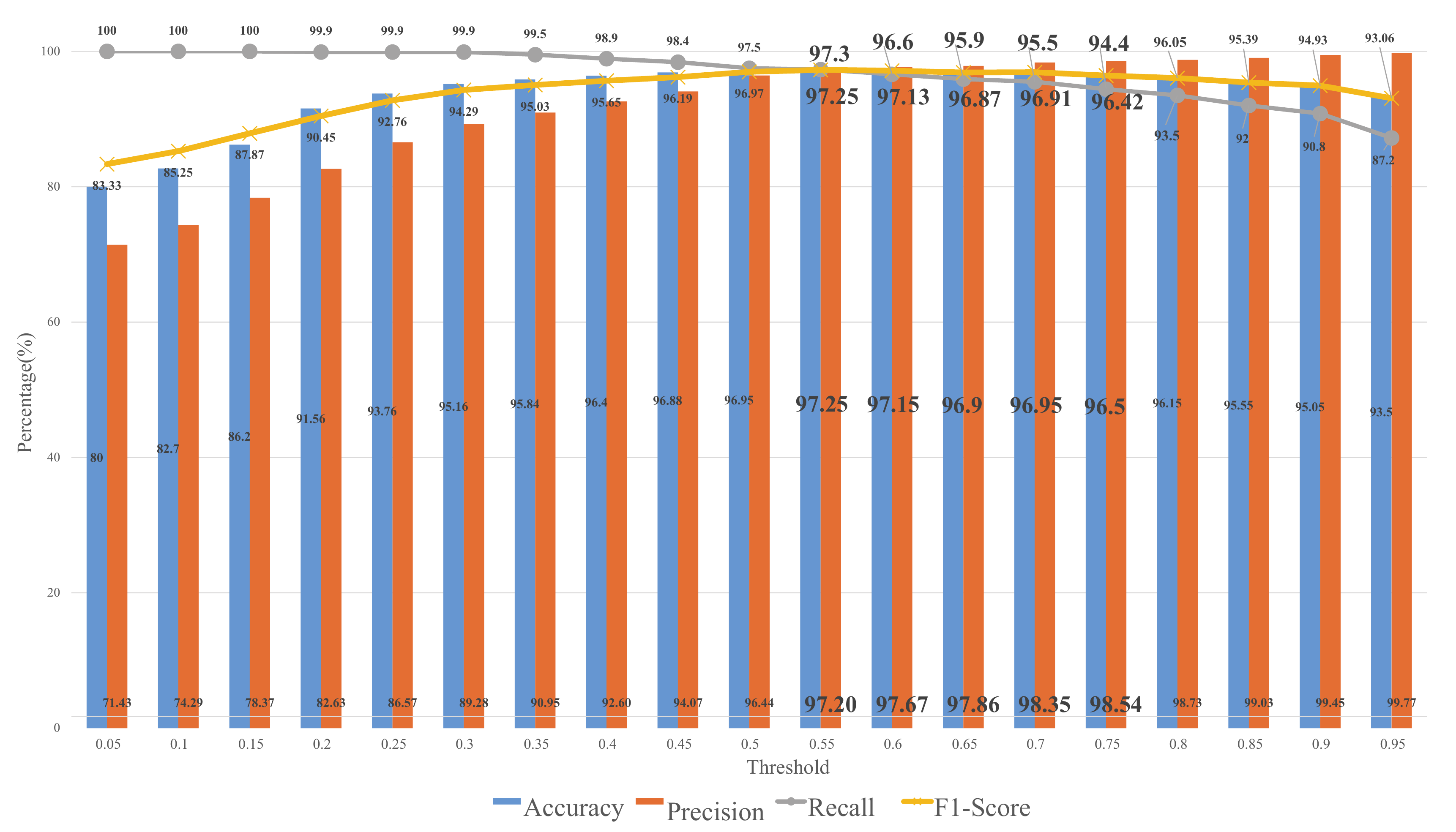}
	\caption{The evaluation results.}\label{fig: F23}
\end{figure}

\subsection{Validation on Real Environment}
To infer the capability of the \textsl{R2-D2} system in the detection of unknown malware, we collected Android apps on Google Play in June 2017. More specifically, we collected Android apps in different categories (e.g., weather, business, finance, travel, etc.) from different countries (e.g., the US, UK, France and Germany, etc.). In addition, we also choose benign samples that none of VirusTotal (https://www.virustotal.com) vendors report malicious. We also selected malicious samples, the criteria for selecting malicious samples is that more than 30 VirusTotal vendors report malicious. We downloaded some malicious samples from contagio (http://contagiominidump.blogspot.tw). In sum, the above sources of benign/malicious samples are used for verifying the detection capability of our \textsl{R2-D2} system. The evaluation metrics in our experiment include True Positive (TP), False Positive (FP), False Negative (FN), True Negative (TN), Accuracy (Acc), Precision (Prec), Recall (Detection Rate, DR), False Positive Rate (FPR) and F1-score (F-measure). The evaluation results are shown in Fig. \ref{fig: F23} with rapid generation and mutation of malware; even the detection model is trained based on our dataset with 1.5 million samples (from January 2017 to May 2017). The precision is higher than accuracy after the threshold 0.6. Moreover, when the threshold is over 0.7, recall and F1-score start to drop, meaning that the detection accuracy is not as good as before. This phenomenon can confirm that though \textsl{R2-D2} is able to reduce the human labor and resource consumption, long-term sample collection and model updating are still necessary. 

\subsection{Real Case study}
We further compare our detection results to the results reported by VirusTotal (Google’s malicious program detection website) and various anti-virus software engines on VirusTotal site. We collected 87 Minecraft apps that are verified by ESET as malware from Google Play. These apps had been active on Google Play since January 2017, and the number of installations have achieved more than 1 million. The most prominent feature of these apps after installation is the extra module for downloading, and the request for “Device Administrator permission”, promoting deceptive advertisements. Google Play has removed these apps. However, if they can be detected in the first place without feature extraction, Android malware can be mitigated. 

From the results shown in Table \ref{tab: tab1}, we can find that most of the above Minecraft malwares can be detected by approximately 16\% vendors (10/60) on 2017/03/24. Until 2017/03/30, approximately 32\% vendors (20/60) can detect these malwares. However, \textsl{R2-D2} can detect more than 75\% of these apps. Our method can not only reduce the resources that are consumed by traditional manual feature extraction method or machine learning method, but also can better detect unknown malware before other malware detection engines can. 

\begin{table}[]
\begin{center}
\caption{The evaluation results of hash.}\label{tab: tab1}
\begin{tabular}{c|c|c|c}
\multirow{2}{*}{} MD5 Hash & \multicolumn{2}{l|}{Date (with VT} & R2-D2 \\ \cline{2-4} 
                  & 03/24 & 03/30 & Result \\ \hline
00e10bedf33cfe4061e45ea632274d53 &     6/61      &     22/59      & 78.78\% \\ \hline
1a72ba660d80564f30d17e3625b999d0 &    32/58      &     34/60      & 99.94\% \\ \hline
ebd3bd60e5b68a52b24f72c3dbe87c0f &    10/60      &     24/59      & 75.29\% \\ \hline
12d2e22f7efdef5950fa82c5331eddde &    37/60      &     40/60      & 99.99\% \\ \hline
84573e568185e25c1916f8fc575a5222 &    23/59      &     25/59      & 98.96\% \\ \hline
e9289ed4df621523372bc1f61e0e5e8c &    10/61      &     22/59      & 75.27\% \\ \hline
8b54fad9660d85442b0732e4ce5bcf8d &    10/61      &     23/58      & 75.29\% \\ \hline
1ceafdb0fc0c7ca1b30f1246e03e1730 &    10/60      &     23/59      & 83.74\% \\ \hline
2b002dc734ecb7caf0c8bb459c7522d9 &    10/60      &     25/59      & 75.28\% \\ \hline
59c83edf9a42b828aa318b36599e7bed &     9/60      &     23/55      & 76.68\%
\end{tabular}
\end{center}
\end{table}

\subsection{Comparison with Existing Methodology}
Finally, we try to compare our \textsl{R2-D2} with DroidSieve \cite{DASPS17} in terms of sample size, DR/FPR, acc, and detection time. From the results shown in Table \ref{tab: tab2}, it is found that \textsl{R2-D2} is slightly weaker than some others in DR/FPR and Acc. This is due to that our \textsl{R2-D2} is trained based on a significantly larger dataset and is more adaptive to the real-world situation. It is well known that different learning-based approaches would perform differently for the difference of the characteristics of its datasets. Others are trained with rather smaller dataset that cannot reflect the real environment, plus they did not provide the data source that they used. Moreover, \textsl{R2-D2} has distinguishing advantage that it has fast detection time, 0.5 seconds for each coming sample. Besides, it does not need to go through manual feature extraction engineering. Asides from that, it takes only 0.4 second for \textsl{R2-D2} to transform an Android apps to color image.

\begin{table}[]
\begin{center}
\caption{The result of comparing R2-D2 and existing research.}\label{tab: tab2}
\begin{tabular}{c|c|c|c|c|c}
Year & Method & Samples & DR/FP (\%) & ACC (\%) & Second \\ \hline
2014 & DenDroid & 1247 & - & 94 & - \\ \hline
2014 & DroidAPIMiner & 3987 & 99/2.2 & - & 15 \\ \hline
2014 & DroidMiner & 2466 & 95.3/0.34 & 92 & 19.8 \\ \hline
2014 & Drebin & 5560 & 94.0/1.0 & - & 0.75 \\ \hline
2014 & DroidSIFT & 2200 & 98.0/5.15 & 93 & 175.8 \\ \hline
2014 & DroidLegacy & 1052 & 93.0/3.0 & 98 & - \\ \hline
2015 & AppAudit & 1005 & 99.3/0.61 & - & 0.6 \\ \hline
2015 & MudFlow & 10552 & 90.1/18.7 & - & - \\ \hline
2015 & Marvin & 15741 & 98.24/0 & - & - \\ \hline
2015 & RevealDroid & 9054 & 98.2/18.7 & 93 & 95.2 \\ \hline
2016 & DroidScribe & 5246 & - & 94 & - \\ \hline
2016 & Madam & 2800 & 96/0.2 & - & - \\ \hline
2017 & DroidSieve & 16141 & 99.3/0 & 99 & 2.5 \\ \hline
2017 & R2-D2 & 829356 & 96/9 & 93 & 0.5
\end{tabular}
\end{center}
\end{table}

\subsection{Large Scale Ransomware Detection with Naked Eye}
Based on our demonstration in OWASP AppSec USA 2017 and RuxCon 2017 \cite{owasp, ruxcon}, here are some new results. We trained a \textsl{R2-D2} detection model with malware samples (including Trojan, ransomware, etc) collected between 2017/1 and 2017/8.
\begin{itemize}
\item Given 20035 malicious samples (collected in Sep. 2017), our model successfully detected 18514 samples among 20035 malicious samples, which implies 91.72\% true positive.
\item Given 20313 benign samples (collected in Sep. 2017), our model unsuccessfully detected 1799 samples among 20313 benign samples, which implies 8.85\% false positive.
\item Given 5852 ransomware samples (collected in Sep. 2017), our model successfully detected 5670 samples among 20313 benign samples, which implies 96.88\% true positive
\end{itemize}
Interestingly, we have found that we can determine ransomware family with naked eye, and the samples from the same ransomware family share similarity in their visual pattern. A fine-grained categorization still needs to rely on our \textsl{R2-D2}, since there are always delicate differences among samples that only \textsl{R2-D2} can identify but human eyes cannot differentiate. Such a similarity in visual pattern of ransomware images helps us do a fast categorization of the ransomware, and reduce labor cost.

\begin{figure}[hbtp]
	\centering
	\includegraphics[width=.85\columnwidth]{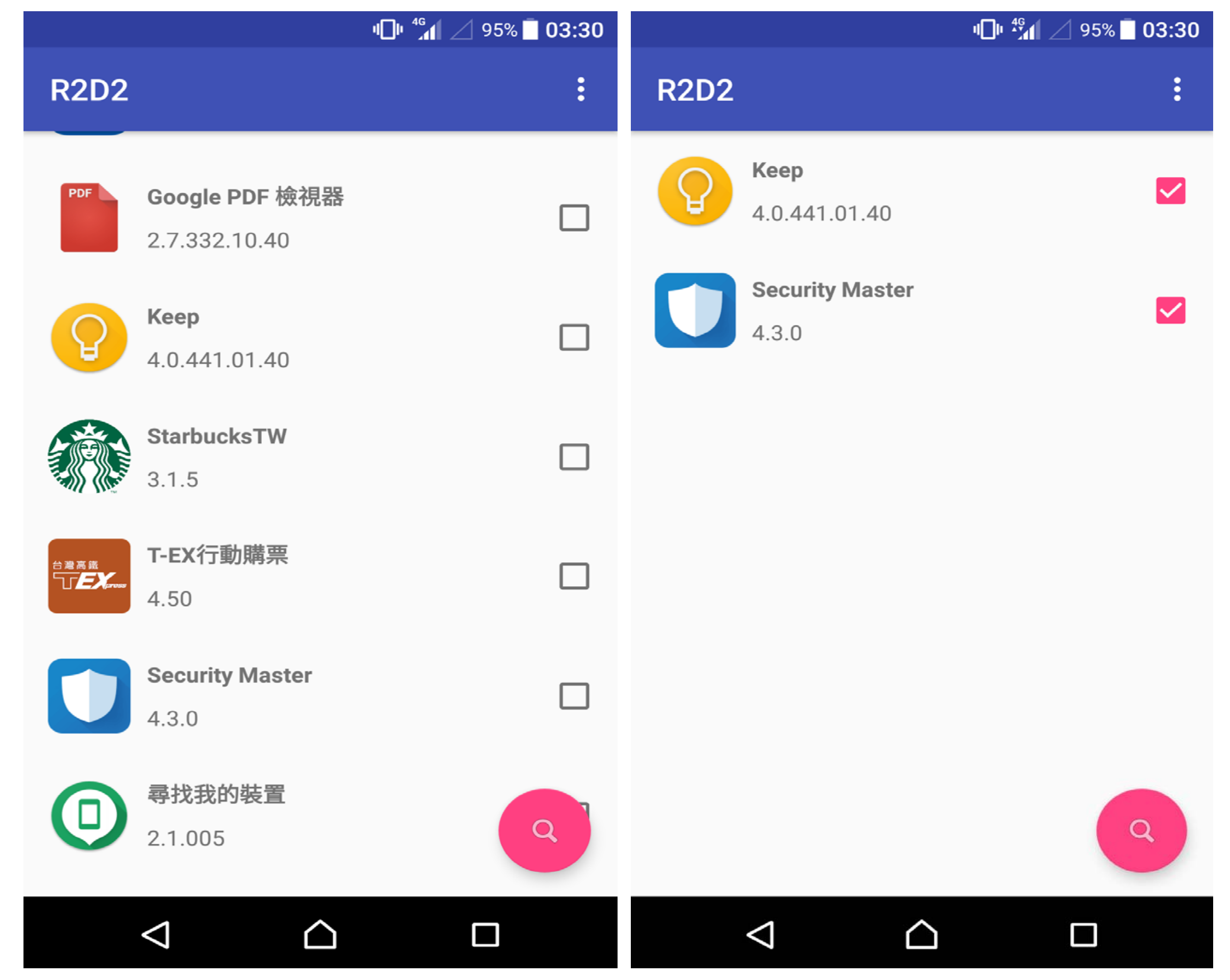}
	\caption{The screenshoot of our \textsl{R2-D2} mobile version.}\label{fig: F29}
\end{figure}

\subsection{Hunting the Ethereum Smart Contract}
Furthermore, the cryptocurrency is attracting more and more attention because of the blockchain technology. Ethereum is gaining a significant popularity in blockchain community, mainly due to the fact that it is designed in a way that enables developers to write smart contracts. However, the security of smart contracts has not received the same attention, and because of its openness, if there is a vulnerability in the contract, anyone can immediately see it, but it cannot be stopped immediately. The DAO vulnerabilities that worth more than \$50 million US dollars is an example \footnote { https://blog.ethereum.org/2016/06/17/critical- update-re-dao-vulnerability, 2016.}. There are many vulnerabilities in the Ethereum smart contract, including "Call Stack Attack \footnote{https://github.com/LeastAuthority/ethereumanalyses/blob/master/\\GasEcon.md\#callstack-depth-limit-errors}", "Time Dependance\footnote{http://martin.swende.se/blog/Breaking the house.html}" and various other known vulnerabilities \cite{surveyvulnerability}. Based on this research, we also translated the bytecode of solidity into RGB color code and then transformed them into a fixed-sized encoded image and fed to our CNN model for training and detecting vulnerability of Ethereum smart contract. With our research, we have demonstrated that the smart contract scan can achieve 0.90+ \cite{AIVILLAGE}.

\section{CONCLUSION}
This research adopts deep learning to construct an end-to-end learning-based Android malware detection method and proposed a color-inspired convolutional neural network (CNN)-based Android Malware detection system, labelled as \textsl{R2-D2}. The results show that our detection system works well in detecting known Android malware and even unknown Android malware. Also, we have integrated the methodology with the back-end of our core product to provide convenient usage scenarios for end users or enterprises. Meanwhile, we have integrated with tensorflow-lite directly into mobile devices (shown in Fig. \ref{fig: F29}). It has made our experimental results proved the effectiveness of our proposed system. The future work is to reduce the complex task and train for higher performance in confronting the Android malware, in order to avoid computation burden.

\section*{Acknowledgement}
This work would not have been possible without the valuable dataset offered by Cheetah Mobile. Special thanks to Dr. Chia-Mu Yu for his support on this research.

\end{document}